\documentclass[letterpaper,conference,%
10pt]{IEEEtran}%
\pdfoutput=1

\usepackage{amsfonts}
\usepackage{amsmath}
\usepackage{amssymb}
\usepackage{graphicx}
\usepackage{amsthm}%

\usepackage{color}
\usepackage{pifont}

\providecommand{\U}[1]{\protect\rule{.1in}{.1in}}
\newtheoremstyle{example}{\topsep}{\topsep}
{}
{}
{\bfseries}
{}
{  }
{\thmname{#1}\thmnumber{ #2}. \thmnote{ (#3)}}
\newtheorem{theorem}{Theorem}

\newtheorem{definition}{Definition}

\newtheorem{lemma}{Lemma}

\theoremstyle{example}

\newcommand{\Tr}{{\rm Tr}}

\newcommand{\nc}{\newcommand}
\nc{\rnc}{\renewcommand}
\nc{\beq}{\begin{equation}}
\nc{\eeq}{{\end{equation}}}
\nc{\beqa}{\begin{eqnarray}}
\nc{\eeqa}{\end{eqnarray}}
\nc{\lbar}[1]{\overline{#1}}
\nc{\bra}[1]{\langle#1|}
\nc{\ket}[1]{|#1\rangle}
\nc{\ketbra}[2]{|#1\rangle\!\langle#2|}
\nc{\braket}[2]{\langle#1|#2\rangle}
\nc{\proj}[1]{| #1\rangle\!\langle #1 |}
\nc{\avg}[1]{\langle#1\rangle}
\nc{\smfrac}[2]{\mbox{$\frac{#1}{#2}$}}
\nc{\tr}{\operatorname{tr}}
\nc{\tracedist}[1]{\Delta_{}\!\left( #1 \right)}
\nc{\fid}[1]{F\!\left( #1 \right)}

\nc{\ox}{\otimes}
\nc{\dg}{\dagger}
\nc{\dn}{\downarrow}
\nc{\cA}{{\cal A}}
\nc{\cB}{{\cal B}}
\nc{\cC}{{\cal C}}
\nc{\cD}{{\cal D}}
\nc{\cE}{{\mathcal E}}
\nc{\cF}{{\cal F}}
\nc{\cG}{{\cal G}}
\nc{\cH}{{\cal H}}
\nc{\cI}{{\cal I}}
\nc{\cJ}{{\cal J}}
\nc{\cK}{{\cal K}}
\nc{\cL}{{\cal L}}
\nc{\cM}{{\cal M}}
\nc{\cN}{{\cal N}}
\nc{\cO}{{\cal O}}
\nc{\cP}{{\cal P}}
\nc{\cR}{{\cal R}}
\nc{\cS}{{\cal S}}
\nc{\cT}{{\cal T}}
\nc{\cU}{{\cal U}}
\nc{\cX}{{\cal X}}
\nc{\cZ}{{\cal Z}}

\nc{\entI}{{\bf I}}
\nc{\entIarrow}{{\bf I}^{\leftarrow}}
\nc{\entH}{{\bf H}}
\nc{\entS}{{\bf S}}
\nc{\entHmin}{H_{\min}}

\nc{\entF}{{\bf E}_f}

\nc{\isom}{\simeq}

\nc{\rank}{\operatorname{rank}}
\nc{\rar}{\rightarrow}
\nc{\lrar}{\longrightarrow}
\nc{\polylog}{\operatorname{polylog}}
\nc{\poly}{\operatorname{poly}}
\nc{\1}{{\openone}}

\nc{\weight}{\textbf{w}}
\nc{\hamdist}{d_{H}}

\def\U{\Upsilon}

\nc{\Sp}{{{\mathbb S}}}
\nc{\RR}{{{\mathbb R}}}
\nc{\CC}{{{\mathbb C}}}
\nc{\FF}{{{\mathbb F}}}
\nc{\NN}{{{\mathbb N}}}
\nc{\ZZ}{{{\mathbb Z}}}
\nc{\PP}{{{\mathbb P}}}
\nc{\QQ}{{{\mathbb Q}}}
\nc{\UU}{{{\mathbb U}}}
\nc{\OO}{{{\mathbb O}}}
\nc{\EE}{{{\mathbb E}}}
\nc{\id}{{\operatorname{id}}}

\nc{\qubitchannel}{\id_2}
\nc{\bitchannel}{\overline{\id}_2}

\nc{\be}{\begin{equation}}
\nc{\ee}{\end{equation}}
\nc{\bea}{\begin{eqnarray}}
\nc{\eea}{\end{eqnarray}}
\nc{\<}{\langle}
\rnc{\>}{\rangle}
\nc{\Hom}[2]{\mbox{Hom}(\CC^{#1},\CC^{#2})}
\nc{\rU}{\mbox{U}}

\nc{\ob}[1]{#1}

\def\mcal{\mathcal}
\def\eps{\epsilon}

\usepackage{ifthen}

\newboolean{WITHAPDX}
\setboolean{WITHAPDX}{true}
\def\ifthen#1#2{\ifthenelse{#1}{#2}{} }

\def\dingone{\ding{172} }
\def\dingtwo{\ding{173} }
\def\dingthree{\ding{174} }
\def\dingfour{\ding{175} }
\def\dingfive{\ding{176} }

\def\rhoFULLatRE{ \rho_{x^{n}\left(  m_{j},l_{j},l_{j-1}\right)  ,x_{1}^{n}\left(l_{j-1}\right)  }^{B_{1\left(  j\right)  }^{n}} }
\def\rhojRE{\rho_{m_j,\ell_j,\ell_{j-1}}^{B_{1(j)}^n }}

\def\rhoRX{\rho^{(j)}_{m_j\ell_j\ell_{j\!-\!1}}\!\!\!\otimes\!\!\!\;\rho^{(j+1)}_{m_{j\!+\!1}\ell_{j\!+\!1}\ell_{\!j}} }
\def\rhojRX{\rho_{m_j,\ell_j,\ell_{j-1}}^{(j)}}
\def\rhojRXmjpr{\rho_{m^\prime_j,\ell_j,\ell_{j-1}}^{(j)}}
\def\rhojBOTH{\rho_{m_j,\ell_j,\ell_{j-1}}^{(j)}}
\def\rhojjRX{\rho_{m_{j+1},\ell_{j+1},\ell_{j}}^{(j+1)}}
\def\rhojjBOTH{\rho_{m_{j+1},\ell_{j+1},\ell_{j}}^{(j+1)}}
\def\rhobarRX{\bar{\rho}_{\ell_j,\ell_{j-1}}^{(j)}}
\def\rhodbarRX{\dbar{\rho}_{|\ell_{j-1}}^{(j)}}

\def\dbar#1{\bar{\bar{#1}}}

\def\PIRElj{\Pi_{\sigma_{\ell_j|\ell_{j-1} } }^{} }
\def\PIREljpr{\Pi_{\sigma_{\ell^\prime_j|\ell_{j-1} } }^{} }

\def\PIREavg{\Pi_{\bar{\sigma}_{|\ell_{j-1}} }^{} }

\def\PIjRXmjlj{\Pi_{\rho_{m_j,\ell_j|\ell_{j-1} } }^{(j)} }
\def\PIjRXmjprlj{\Pi_{\rho_{m^\prime_j,\ell_j|\ell_{j-1} } }^{(j)} }
\def\PIjRXmjprljpr{\Pi_{\rho_{m^\prime_j,\ell^\prime_j|\ell_{j-1} } }^{(j)} }
\def\PIjRXlj{\Pi_{\bar{\rho}_{\ell_j|\ell_{j-1} } }^{(j)} }
\def\PIjRXljpr{\Pi_{\bar{\rho}_{\ell^\prime_j|\ell_{j-1} } }^{(j)} }
\def\PIjRXavg{\Pi_{\dbar{\rho}_{|\ell_{j-1} } }^{(j)} }
\def\PIjjRXlj{\Pi_{\tau_{\ell_j } }^{(j+1)} }
\def\PIjjRXljpr{\Pi_{\tau_{\ell^\prime_j } }^{(j+1)} }
\def\PIjjRXavg{\Pi_{\bar{\tau}  }^{(j+1)} }

\def\PRE{P_{\ell_j|\ell_{j-1}}^{B^n_{1(j)} } }
\def\PREpr{P_{\ell_j^\prime|\ell_{j-1}}^{B^n_{1(j)} } }
\def\PRX{P_{m_j,\ell_j|\ell_{j-1}}^{B^n_{(j)} B^n_{(j+1)} } }

\def\PRXljmjpr{P_{m^\prime_j,\ell_j|\ell_{j-1}}^{B^n_{(j)} B^n_{(j+1)} } }
\def\PRXljprmjpr{P_{m^\prime_j,\ell^\prime_j|\ell_{j-1}}^{B^n_{(j)} B^n_{(j+1)} } }

\def\PRXj{P_{m_j,\ell_j|\ell_{j-1}  }^{B^n_{(j)}} }
\def\PRXjmjpr{P_{m^\prime_j,\ell_j|\ell_{j-1}  }^{B^n_{(j)}} }
\def\PRXjljprmjpr{P_{m^\prime_j,\ell^\prime_j|\ell_{j-1}  }^{B^n_{(j)}} }
\def\PRXjj{P_{\ell_j|\ell_{j-1}  }^{B^n_{(j+1)}} }
\def\PRXjjpr{P_{\ell^\prime_j|\ell_{j-1}  }^{B^n_{(j+1)}} }

\def\GAMRE{\Gamma_{\ell_j|\ell_{j-1}  }^{B^n_{1(j)} } }
\def\LAMRX{\Lambda_{m_j,\ell_j|\ell_{j-1}  }^{B^n_{(j)}B^n_{(j+1)} } }

\def\ExpXgUXone{ \mathop{\mathbb{E}}_{X^n|U^n\!X_1^n } }
\def\ExpUXone{ \mathop{\mathbb{E}}_{U^n\!X_1^n } }
\def\ExpUXgXone{ \mathop{\mathbb{E}}_{U^n\!X^n|X_1^n } }
\def\ExpALL{ \mathop{\mathbb{E}}_{U^n\!X^n\!X_1^n } }

\def\ExpXone{ \mathop{\mathbb{E}}_{X_1^n} }

\begin{document}

	\title{Partial decode-forward for quantum relay channels} 

	\author{
		\IEEEauthorblockN{
			Ivan Savov\IEEEauthorrefmark{1}, 
			Mark M. Wilde\IEEEauthorrefmark{1} and 
			Mai Vu\IEEEauthorrefmark{2} 
		} 
	\IEEEauthorblockA{
		\!\!\!\IEEEauthorrefmark{1}
		School of Computer Science and 
		\IEEEauthorrefmark{2}
		Electrical and Computer Engineering Department, McGill University,  \textit{Montr\'eal, Canada} \\
	} 
}

	\maketitle
	
	\vspace*{-5mm}
	\begin{abstract}

	A relay channel is one in which a Source and Destination
	  use an intermediate Relay station in order to improve communication rates.
	  We propose the study of relay channels
	  with classical inputs and quantum outputs
	  and prove that  %
	a ``partial decode and forward'' strategy is achievable.
	We divide the channel uses into many blocks and build codes
	   in a randomized, block-Markov manner within each block. 
	The Relay performs a standard Holevo-Schumacher-Westmoreland
	  quantum measurement on each block in order to decode part of the Source's message and then
	  forwards this partial message in the next block. 
	The Destination performs a novel ``sliding-window'' quantum measurement on
	  two adjacent blocks in order to decode %
	  the Source's message. 
	  This strategy achieves non-trivial rates for classical communication over a quantum relay channel.

	\end{abstract}

\section{Introduction}

	Suppose that a Source wishes to communicate with a remote Destination.
	Suppose further that a Relay is available that can decode the messages 
	transmitted by the Source during one time slot and \emph{forward} them to the Destination
	during the next time slot.
	With the Relay's help, the Source and Destination can improve communication rates
	because the Destination can decode the intended messages in parallel from the channel outputs at 
	two consecutive time slots.
	In this way, useful information is received both from the Source and the Relay.

	The relay channel has been studied extensively in the context of
	classical information theory \cite{cover1979capacity, %
	xie2005achievable, el2010lecture}. 
	There, the discrete memoryless
	relay channel %
	is modelled as a conditional probability 
	distribution $p(y_1,y|x,x_1)$, where $y_1$ and $y$ are the
	respective outputs at the Relay and Destination whenever the Source and
	Relay input symbols $x$ and $x_1$.
	Two important families of coding strategies exist for relay channels: 
	compress-and-forward and decode-and-forward \cite{cover1979capacity,el2010lecture}.
	The partial decode-and-forward strategy differs from the decode-and-forward 
	strategy in that it has the Relay decode only
	\emph{part} of the message from the Source \cite{cover1979capacity}. %

	The study of quantum channels with information-theoretic techniques
	has been an active area for some time now \cite{wilde2011book}. %
	Theoretical interest has focused on classical-quantum channels 
	of the form $(\mcal{X}, \ \mcal{N}^{X\to B}(x)\!\equiv\! \rho^B_x, \ \mcal{H}^B)$, 
	where, for each of the inputs $x \in \mcal{X}$,
	there corresponds an output quantum state,
	described by a density operator $\rho^B_x$
	in a finite-dimensional Hilbert space $\mcal{H}^B$.
	Classical-quantum channels are a useful abstraction for studying general 
	quantum channels and correspond to the transmitters being restricted to classical encodings.		
	In this setting, single-letter formulas characterize the 
	capacity of point-to-point \cite{H98,SW97} and
	multiple-access channels \cite{winter2001capacity} 
	and give achievable rates for other network channels 
	\cite{FHSSW11,S11a,SW11}.
	The study of quantum channels finds practical applications in optical communications.
	Bosonic channels model the quantum aspects of optical communication channels,
	where information is encoded into continuous degrees of freedom.
	It is known that collective quantum measurements on bosonic-channel outputs
	outperform classical strategies,
	particularly in the low-photon-number regime \cite{PhysRevLett.92.027902}.
	In other words, quantum measurements are \emph{necessary} to
	achieve their
	ultimate capacity. %
	Ref.~\cite{PhysRevLett.92.027902} also demonstrates that
	classical encoding is \emph{sufficient} 
	to achieve the Holevo capacity of the lossy bosonic channel,
	giving further motivation for the theoretical study of classical-quantum models.
	 In this paper, we develop a ``partial decode and forward'' strategy 
	for classical-quantum relay channels. 
	Our results here are the first extension of the quantum
	 simultaneous decoding techniques used in \cite{FHSSW11,S11a} 
	to multi-hop networks. 
	In the partial-decode-and-forward strategy given here, the Relay decodes part of the Source's
	 message in one block and forwards it in the next.
	 The Destination performs a novel ``sliding-window'' quantum measurement to decode
	both parts of the Source's message in two consecutive blocks \cite{carleial1982multiple,xie2005achievable} and
	 in doing so allows for the Source and Destination to achieve non-trivial communication rates.
	We state our main result in the Section~\ref{sec:results},
	introduce the necessary background
	on quantum systems and quantum decoding in Section~\ref{sec:preliminaries},
	and give the proof in Section~\ref{sec:ach-proof}.
	We conclude and discuss open problems in Section~\ref{sec:discussion}.

\section{Results}
	\label{sec:results}

	A classical-quantum relay channel $\mcal{N}$ is a map with two classical inputs
	$x$ and $x_1$ and two output quantum systems $B_1$ and $B$.
	For each pair of possible input symbols $(x,x_1)\in \mcal{X} \times \mcal{X}_1$, the channel prepares
	a density operator $\rho^{B_1B}_{x,x_1}$ defined on the
	 tensor-product Hilbert space $\mcal{H}^{B_1}\otimes \mcal{H}^{B}$:
	\be
		\rho^{B_1B}_{x,x_1} \equiv   \mcal{N}^{XX_1\to B_1B}(x,x_1),
		\label{eqn:relay-chan-def}
	\ee
	 where $B_1$ is the Relay output and
	$B$ is the Destination output.

    The theorem below captures the main result of our paper:
	\begin{theorem}[Partial decode-forward inner bound]		\label{thm:main}
		Let $\{ \rho_{x,x_1} \}$ be a cc-qq relay channel as in \eqref{eqn:relay-chan-def}.
		Then a rate $R$ is achievable, provided that
		the following inequality holds:
		\be
		R\leq\max_{
		p\left(  u,x,x_{1}\right)  }\min\left\{  
		{ 
		I\!\left(  XX_{1};B\right)  _{\theta},
		\atop
		I\!\left( U;B_{1}|X_{1}\right)  _{\theta}+I\!\left(  X;B|X_{1}U\right)  _{\theta}
		}
		\right\},
		\ee
		where the information quantities are with respect to the %
		classical-quantum state $\theta^{UXX_{1}B_{1}B}\equiv$%
		\begin{equation}
		\sum_{x,u,x_{1}}
		p\left(  u,x,x_{1}\right)
		\left\vert u\right\rangle\!\!
		\left\langle u\right\vert ^{U}\otimes\left\vert x\right\rangle\!\!\left\langle
		x\right\vert ^{X}\otimes\left\vert x_{1}\right\rangle\!\!\left\langle
		x_{1}\right\vert ^{X_{1}}\otimes\rho_{x,x_{1}}^{B_{1}B}.\label{eq:code-state}%
		\end{equation}
	\end{theorem}
	
	Our code construction employs codebooks 
	$\{ x_1^n \}$,
	$\{ u^n \}$, and 	$\{ x^n \}$
	generated according to the distribution $p(x_1)p(u|x_1)p(x|u,x_1)$.
	We split the message for each block into two parts 
	$%
	(m,\ell) \in \mathcal{M}\times\mathcal{L}$
	such that the rate $R = R_m + R_\ell$.
	The Relay fully decodes the message $\ell$ and re-encodes it
	directly in the next block (without using binning).
	The Destination exploits a ``sliding-window'' decoding strategy
	\cite{carleial1982multiple,xie2005achievable} 
	by performing a collective measurement on two consecutive blocks. In this approach,
	the message pair $(m_j,\ell_j)$ sent during block $j$ is decoded 
	from the outputs of blocks $j$ and $j+1$,
	using an  ``{\sc and}-measurement.''

\section{Preliminaries}
	\label{sec:preliminaries}

	In this section, we introduce the notation used in our paper and
	some background information on quantum decoding.

	\subsubsection{Quantum systems}

		We denote quantum systems as $B_1$ and $B$ and the corresponding Hilbert
		spaces as $\mathcal{H}^{B_1}$ and $\mathcal{H}^{B}$.
		We represent quantum states of a system~$B$ with a density operator $\rho^{B}$,
		which is a positive semi-definite operator with unit trace.
		Let $H(B)_{\rho}\equiv-\text{Tr}\left[  \rho^{B}\log_2\rho^{B}\right]$ 
		denote the von Neumann entropy of the state $\rho^{B}$. 
	In order to describe the ``distance'' between two quantum states, we use the notion
	of \emph{trace distance}.
	The trace distance between states $\sigma$ and $\rho$ is
	$\|\sigma-\rho\|_1 = \mathrm{Tr}|\sigma - \rho|$, where  $|X| = \sqrt{X^{\dagger}X}$ \cite{wilde2011book}.
	Two states can \emph{substitute} for one
	another up to a penalty proportional to the trace distance between them:
	\begin{lemma} \label{lem:tr-trick}
	Let 	
	$0\leq \rho, \sigma, \Lambda \leq I$. Then
	\be
	\mathrm{Tr}\left[  \Lambda\rho\right]
		\leq
	\mathrm{Tr}\left[ \Lambda \sigma\right]  
		+ \left\Vert \rho-\sigma\right\Vert _{1}.
	\label{eqn:tr-trick}
	\ee
	\end{lemma}
	\begin{IEEEproof}
	This follows from a variational characterization of 
	trace distance as the distinguishability of
	the states under an optimal measurement $M$ \cite{wilde2011book}:
	$\left\Vert \rho-\sigma\right\Vert _{1} = 2
	\max_{0 \leq M \leq I} \mathrm{Tr}\left[ M(\rho-\sigma) \right]$.
	\end{IEEEproof}

	\medskip

	\subsubsection{Quantum decoding}

		In a communication scenario, the decoding operations performed by the 
		receivers correspond to quantum measurements on the outputs of the channel.
		A quantum measurement 	is a positive operator-valued measure (POVM) 
		$\left\{ \Lambda_{m}\right\}_{m\in \mcal{M}}$ %
		on
		the system $B^n$. %
		To be a valid POVM, the set $\{\Lambda_{m}\}$ of 
		$|\mcal{M}|$ operators should all be positive semi-definite
		and sum to the identity:  $\Lambda_{m} \geq 0, \,\,\, \sum_{m}\Lambda_{m}=I$.

		Suppose we are given positive operators $\{ P_m \}_{m\in\mcal{M}}$ 
		that are  apt at detecting ($\text{Tr}\!\left[P_m\;\rho_m\right] \geq 1- \epsilon$)
		and distinguishing  ($\text{Tr}\!\left[P_{m}\;\rho_{m^\prime\neq m}\right] \leq \epsilon$)
		the output states produced by each message.
		We can construct a valid POVM (known as the square-root
		measurement \cite{H98,SW97}) by \emph{normalizing} 
		these operators:
		\begin{align}
		\Lambda_{m} &  \equiv 
		\left(  
			\sum_{k}P_{k}
		\right)^{\!\!\!-1/2}
		\!\! P_{m}
		\left(
			\sum_{k} P_{k}
		\right)^{\!\!-1/2}\!\!. \label{eq:square-root-POVM-generic} 
		\end{align}
		The error analysis of a square-root measurement is  
		greatly simplified by using the Hayashi-Nagaoka operator inequality.
		
		\begin{lemma}[Hayashi-Nagaoka \cite{hayashi2003general}] \label{lem:HN-inequality}
		If $S$ and $T$ are operators such that  
		$0\leq T$ and $0\leq S\leq I$, then
		\be
			I-\left(  S+T\right)  ^{-\frac{1}{2}}S\left(  S+T\right)  ^{-\frac{1}{2}}%
			\ \: \leq \ \:
			2\left(  I-S\right) \  + \ 4T.
		\ee
		\end{lemma}%
		\noindent

	\medskip

	\subsubsection{Error analysis}
		In the context of our coding strategy,
		we analyze the average probability of error at the Relay:
		\begin{align*}
			\bar{p}_{e}^{R}%
			\equiv 
				\frac{1}{|\mathcal{L}|}\sum_{\ell_j}					
				\text{Tr}\!
				\left\{  
					\left(  I-\Gamma^{B_{1(j)}^{n}} _{\ell_j} \right)
					\rho_{\ell_j }^{B_{1(j)}^{n}} 
				\right\},
		\end{align*}
		and the average probability of error %
		at the Destination:
		\begin{align}
			\label{eqn:avgpe-De}
			\bar{p}^{D}_{e}%
			\!  \equiv \! 
				\frac{1}{|\mathcal{M}||\mathcal{L}|}\sum_{m_j,\ell_j}
				\text{Tr}\!
				\left[ \! 
					\left(\!  I-\Lambda_{m_j,\ell_j}^{B_{(j)}^{n} B_{(j+1)}^{n} }  \right)
					\rho_{m_j,\ell_j  }^{B_{(j)}^{n} B_{(j+1)}^{n} } 
				\right].
		\end{align}
		The operators $\left(  I-\Gamma_{\ell_j} \right)$ 
		and $\left(  I-\Lambda_{m_j,\ell_j}\right)$
		correspond to the complements of the correct decoding outcomes.

		    \begin{definition}%
			An $(n,R,\epsilon)$ partial-decode-and-forward code for the quantum relay channel consists
			of two codebooks %
			$\{x^n(m_j,\ell_j)\}_{m_j\in \mathcal{M}, \ell_j \in \mathcal{L}}$ and
			$\{x_1^n(\ell_j)\}_{\ell_j \in \mathcal{L}}$
			and decoding POVMs  
			$\left\{ \Gamma_{\ell_j}\right\}_{\ell_j\in \mathcal{L}}$ 
			and 
			$\left\{  \Lambda_{m_j,\ell_j}\right\}_{m_j\in \mathcal{M}, \ell_j \in \mathcal{L}}$
			such that the average probability of error
			is bounded from above as 
			$\overline{p}_{e} = \bar{p}^{R}_{e} + \bar{p}^{D}_{e} \leq \epsilon$.
		    \end{definition}  
		
		A rate $R$ is \textit{achievable} if there exists
		an $\left(  n,R-\delta,\epsilon\right)$ quantum relay channel code
		for all $\epsilon,\delta>0$ and sufficiently large $n$.

\section{Achievability proof}
	\label{sec:ach-proof}

	The channel is used for $b$ blocks, each indexed by $j \in \{1,\ldots,b\}$.
	Our error analysis shows that:
	\begin{itemize}

	\item 
	The Relay can decode  the message $\ell_{j}$ during  block $j$.

	\item 
	The Destination can simultaneously decode $(m_j,\ell_j)$ 
	from a collective measurement on the output systems 
	of blocks $j$  and $j+1$.
	
	\end{itemize}

	The error analysis at the Relay is similar
	to that of the Holevo-Schumacher-Westmoreland theorem \cite{H98,SW97}.
	The message $\ell_{j}$ can
	be decoded reliably, %
	if the rate $R_{\ell}$ %
	obeys the following inequality:
	\be
	R_{\ell}\leq I\left(  U;B_{1}|X_{1}\right) _{\theta}.
	\label{eqn:bound-from-relay}
	\ee
	\ifthenelse{\boolean{WITHAPDX}}{
		We give a proof in the Appendix.
	}{
		The error analysis %
		has been omitted for brevity.
	}

	The decoding at the Destination is a variant of the quantum simultaneous  
	decoder from \cite{FHSSW11,S11a}.
	To decode the message $(m_j,\ell_j)$, the Destination performs a
	``sliding-window'' decoder, implemented 
	as an ``{\sc and}-measurement'' on the outputs of blocks $j$ and $j+1$.
	This coding technique does not require binning at the Relay or backwards decoding
	at the Destination \cite{carleial1982multiple,xie2005achievable}.
	
	In this section, we give the details of the coding strategy
	and analyze the probability of error at the Destination.

\medskip

\textbf{Codebook construction}. 
Fix a distribution $p(u,x,x_1)$ and
 independently generate a different codebook for each block $j$:
\begin{itemize}
\item 
Randomly and independently generate $2^{nR_{\ell}}$
sequences $x_{1}^{n}\!\left(  \ell_{j-1}\right)$, $\ell_{j-1} \in\left[1: 2^{nR_{\ell}}\right]$,
according to  $\prod\limits_{i=1}^{n}p%
\!\left(  x_{1i}\right)$.

\item 
For each $x_{1}^{n}\!\left(  \ell_{j-1}\right)$, 
randomly and conditionally independently generate 
$2^{nR_{\ell}}$  sequences $u^{n}\!\left(  \ell_{j}|\ell_{j-1}\right)$,
$\ell_{j} \in\left[1: 2^{nR_{\ell}}\right]$
according to %
$\prod\limits_{i=1}^{n}p%
\left(  u_{i}|x_{1i}\!\left(  \ell_{j-1}\right)
\right)  $. 

\item
For each $x_{1}^{n}\!\left(  \ell_{j-1}\right)$
and each corresponding $u^{n}\!\left(  \ell_{j}|\ell_{j-1}\right)$,
randomly and conditionally independently generate
$2^{nR_{m}}$  sequences 
$x^{n}\!\left(  m_{j}|\ell_{j},\ell_{j-1}\right) $, $m_{j} \in\left[1: 2^{nR_{m}}\right]$,
according to the distribution:
$\prod\limits_{i=1}^{n}
p%
\big(  x_{i}  | x_{1i}\!\left(  \ell_{j-1}\right)  ,u_{i}\!\left(\ell_{j}|\ell_{j-1}\right)  \big)$.

\end{itemize}

\medskip
\textbf{Transmission}. The transmission of $(\ell_{j},m_{j})$ to the Destination
happens during blocks $j$ and $j+1$. At the beginning of block $j$, we assume
that the Relay has correctly decoded the message $\ell_{j-1}$. During
block $j$, the Source inputs the new messages $m_{j}$ and $\ell_{j}$, and the
Relay forwards the old message $\ell_{j-1}$. That is, their inputs to the channel
for block $j$ are the codewords $x^{n}\!\left(  m_{j},\ell_{j},\ell_{j-1}\right)  $
and $x_{1}^{n}\!\!\left(  \ell_{j-1}\right)  $, leading to the following state at the
channel outputs:%
\[
\rhojBOTH \equiv
\rho_{x^{n}\left(  m_{j},\ell_{j},\ell_{j-1}\right)  ,x_{1}^{n}\left(
\ell_{j-1}\right)  }^{B_{1\left(  j\right)  }^{n}B_{\left(  j\right)  }^{n}}.
\]

During block $j+\!1$, the Source transmits $(m_{j+1},\ell_{j+1})$ given $\ell_j$, whereas the
Relay sends $\ell_{j}$, leading to the state:
\[
\rhojjBOTH \equiv 
\rho_{x^{n}\left(  m_{j+1},\ell_{j+1},\ell_{j}\right)  ,x_{1}^{n}\left(
\ell_{j}\right)  }^{B_{1\left(  j+1\right)  }^{n}B_{\left(  j+1\right)  }^{n}}.
\]
Our shorthand notation is such that the states are identified
by the messages that they encode, and the codewords are
implicit.

	\medskip
	
	\textbf{Decoding at the Destination}. 
	We now determine a decoding POVM\ that the Destination can perform on the output
	systems spanning blocks $j$ and $j+1$. 
	The Destination is trying to recover messages $\ell_{j}$ and $m_{j}$ given knowledge
	of $\ell_{j-1}$.

	First let us consider forming decoding operators for block $j+1$.
	Consider the state obtained by tracing over the systems $X$, $U$, and $B_{1}$
	in (\ref{eq:code-state}):%
	\[
	\theta^{X_1B} = 
	\sum_{x_{1}}p\left(  x_{1}\right)  \left\vert x_{1}\right\rangle\!\! \left\langle
	x_{1}\right\vert ^{X_{1}}\otimes \tau_{x_{1}}^{B},
	\]
	where $
	\tau_{x_{1}}^{B}\equiv\sum_{u,x}p\left(  u|x_{1}\right)  p\left(
	x|x_{1},u\right)  \rho_{x,x_{1}}^{B}$.
	Also, let $\bar{\tau}^{B}$ denote the following state:
	$
	\bar{\tau}^{B}\equiv\sum_{x_{1}}p\left(  x_{1}\right)  \tau_{x_{1}}^{B}.
	$
	Corresponding to the above states are conditionally typical projectors \cite{wilde2011book} of the
	following form:%
	\begin{align*}
	\PIjjRXlj & \equiv
	\Pi_{\tau_{x_{1}^{n}\left(  \ell_{j}\right)  }}^{B_{\left(  j+1\right)  }^{n}}, \qquad %
	\PIjjRXavg   \equiv \Pi_{\bar{\tau}}^{B_{\left(  j+1\right)  }^{n}},
	\end{align*}
	which we combine to form the positive operator:
	\begin{align}
	\PRXjj
	&\equiv
	\PIjjRXavg \ \PIjjRXlj \ \PIjjRXavg, \label{eqn:PRX-def-II}
	\end{align}		
	that acts on the output systems $B^{n}_{(j+1)}$ of block $j+1$.

	\smallskip

	Let us now form decoding operators for block $j$. 
	Define the conditional typical
	projector for the state
	$\rhojRX$ 
	as%
	\be
	\PIjRXmjlj
	\equiv \ 
	\Pi_{\rho_{x^{n}\left(  m_{j},l_{j},l_{j-1}\right)  ,x_{1}^{n}\left(
	l_{j-1}\right)  }}^{B_{\left(  j\right)  }^{n}}.
	\ee	
	The state obtained from
	\eqref{eq:code-state} by tracing over $X$ and $B_{1}$ is
	\[
	\theta^{UX_1B}=
	\sum_{u,x_{1}}p\left(  u|x_{1}\right)  p\left(  x_{1}\right)  \left\vert
	u\right\rangle\!\! \left\langle u\right\vert ^{U}\otimes\left\vert x_{1}
	\right\rangle \!\!\left\langle x_{1}\right\vert ^{X_{1}}\otimes\bar{\rho}_{u,x_{1}}^{B},
	\]
	where $
	\bar{\rho}_{u,x_{1}}^{B}\equiv\sum_{x}p\left(  x|x_{1},u\right)  \rho_{x,x_{1}}^{B}
	$.
	Define also the doubly averaged state
	$
		\dbar{\rho}_{x_{1}}^{B}\equiv\sum_{u,x}p\left(  x|x_{1},u\right)p\left(  u|x_{1}\right)  \rho_{x,x_{1}}^{B}.
	$

	The following conditionally typical projectors  will be
	useful in our decoding scheme:%
	\begin{align*}
	\PIjRXlj 
	& \equiv
	 \Pi_{\bar{\rho}_{u^{n}\left(  l_{j},l_{j-1}\right)  ,x_{1}^{n}\left(
	l_{j-1}\right)  }}^{B_{\left(  j\right)  }^{n}},\,\,\,\,
	\PIjRXavg
	& \equiv
	\  \Pi_{\dbar{\rho}_{x_{1}^{n}\left(  l_{j-1}\right)  }}^{B_{\left(  j\right)  }^{n}%
	}.
	\end{align*}
	We can then form a positive operator 
	``sandwich'':
	\begin{align}
	\!\!
	\PRXj 
	\!\!\!   &\equiv 
	\PIjRXavg \PIjRXlj \PIjRXmjlj \!\!\PIjRXlj \!\!\!\PIjRXavg\!\!\!. \label{eqn:PRX-def-I} 
	\end{align}
	Finally, we combine the positive operators from 
	\eqref{eqn:PRX-def-II} and \eqref{eqn:PRX-def-I} to form the
	``sliding-window'' positive operator:
	\be
		\PRX=\PRXj \otimes \PRXjj,
	\ee			
	from which we can build the Destination's square-root measurement 
	$\LAMRX$ using the formula in  \eqref{eq:square-root-POVM-generic}. This measurement is what
	 we call the ``{\sc and}-measurement.''

\medskip
\textbf{Error analysis at the Destination}. 
	In this section, we prove that the Destination can correctly 
	decode the message pair $(m_j,\ell_j)$ by employing the measurement 
	$\{\LAMRX\}$ on the output state $\rhoRX$ spanning blocks $j$ and $j+1$.
	The %
	average probability of error for the Destination is given
	in \eqref{eqn:avgpe-De}.
	 For now, we consider
	the error analysis for a single message pair $(m_j,\ell_j)$:
	\begin{align*}
	 \bar{p}_e^{D}
	 & \equiv
	\!\text{Tr}\!\left[ 
		\left(I   - \!\LAMRX \right)  
		\rhoRX
	  \right]. \\
	& \leq  2\ 
	\text{Tr}\left\{  \left(  I %
	-\PRX
	\right)  \ \ 
	\rhoRX
	\right\}  
	\\
	&  \quad +4 \!\!\!\!\!\!\!\!\!\! 
	\sum_{\left(  \ell_{j}^{\prime},m_{j}^{\prime}\right)  \neq\left(\ell_{j},m_{j}\right)  }
	\!\!\!\!\!\!\!
	\text{Tr}
	\left\{  
	\PRXljprmjpr \ \
	\rhoRX
	\right\}\!,
	\end{align*}
	where use the Hayashi-Nagaoka inequality (Lemma~\ref{lem:HN-inequality})
	to decompose the error operator $(I-\LAMRX)$ into two
	components:
	(I) a term corresponding to the probability that the correct detector
	 does not ``click'': $(  I-\PRX )$,
	and (II) another term corresponding to the probability that a wrong detector ``clicks'': 
	$\sum_{\left(  \ell_{j}^{\prime},m_{j}^{\prime}\right)} \PRXljprmjpr$.
	These two errors are analogous to the classical error events 
	in which an output sequence $y^n$ is either not jointly typical with the correct codeword
	 or is jointly typical with another codeword.

	We will bound the expectation %
	of the average probability of error $\ExpALL\!\!\left\{ \bar{p}_e^{D} \right\}$,
	using the properties of typical projectors \cite{wilde2011book},
	and the following %
	lemmas:
	\begin{lemma} 
		\label{lem:operator-union-bound}
		For any operators  $0 \leq P^A, Q^B \leq I$, we have:
		\[
			(I^{AB} - P^A\!\otimes\!Q^B) \leq (I^{A}\! -\! P^{A})\!\otimes\! I^{B} + I^{A}\!\otimes\!(I^{B} - Q^{B}).
		\]
	\end{lemma}
	
	\begin{IEEEproof}
	Expand and rearrange 	$(I-P)\otimes(I-Q) \geq 0$.
	\end{IEEEproof}

	\begin{lemma}[Gentle Operator Lemma for Ensembles \cite{itit1999winter}] 
	Let $\left\{  p\!\left(  x\right)  ,\rho_{x}\right\}$ be an ensemble
	and let $\bar{\rho} \equiv\sum_{x}p\!\left(  x\right)  \rho_{x}$. %
	If an operator $\Lambda$, where $0 \leq \Lambda \leq I$, 
	has high overlap with the average state, %
	$\mathrm{Tr}\left[  \: \Lambda \: \bar{\rho} \: \right]  \geq1-\epsilon$,
	then the subnormalized state $\sqrt{\Lambda}\rho_{x}\sqrt{\Lambda}$ is close in 
	trace distance to the original state $\rho_{x}$ on average: %
	$ 
	\mathbb{E}_{X}\left\{  \left\Vert \sqrt{\Lambda}\rho_{X}\sqrt{\Lambda}%
	-\rho_{X}\right\Vert _{1}\right\}  \leq2\sqrt{\epsilon}.
	$
	\label{lem:gentle-operator}
	\end{lemma}

	The first term (I) is bounded as follows:
	\begin{align*}
	&\text{Tr}\!\left[  
		\left(  I-\PRX \right)  \ 
		\rhoRX  
	\right]  
	\\
	& =
		\text{Tr}\!\left[  
			\left(  I-\PRXj \!\!\!\otimes \!\PRXjj \right) 
			\rhoRX  
		\right]  \\
	& \leq
		\underbrace{
		\text{Tr}\!\left[  
			\left(  I-\PRXj  \right) 
			\rhojRX 
		\right] 
		}_{\alpha}
		\underbrace{
		\text{Tr}\!\left[  
			\rhojjRX
		\right]
		}_{=1}	
		 \\
		& \quad + 
		\underbrace{
		\text{Tr}\!\left[  
			\rhojRX 
		\right] 
		}_{=1}
		\underbrace{			
		\text{Tr}\!\left[  
			\left(  I- \PRXjj \right) 
			\rhojjRX  
		\right]
		}_\beta,
	\end{align*}%
	where the inequality follows from Lemma~\ref{lem:operator-union-bound}.

We proceed to bound the term $\beta$ as follows:
\begin{align*}
\beta 
&=	
	\text{Tr}\!\left[  
		\left(  I- \PRXjj \right) 
		\rhojjRX  
	\right]  \\
&=
	\text{Tr}\!\left[  
		\left(  I- \PIjjRXavg \PIjjRXlj  \PIjjRXavg \right) 
		\rhojjRX  
	\right]  \\
&=
	1 - 
	\text{Tr}\!\left[  
		\PIjjRXavg \PIjjRXlj  \PIjjRXavg \ 
		\rhojjRX  
	\right]  \\
&\leq
	1 -
	\text{Tr}\!\left[  
	  \PIjjRXlj \ 
		\rhojjRX  
	\right]  \\
&\qquad \quad 
  +  \left\| \PIjjRXavg \rhojjRX \PIjjRXavg   - \rhojjRX \right\|_1,
\end{align*}
where the inequality follows from Lemma~\ref{lem:tr-trick}.

By taking the expectation over the code randomness,
we obtain the upper bound:
\begin{align*}
 \hspace{-2mm}\ExpALL\!\!\!\left\{ \beta \right\}
&=
	1 - 
	\ExpXone
	\text{Tr}\!\left[  
	  \PIjjRXlj \! 
		\ExpUXgXone \!\!\!\left\{ \rhojjRX   \right\}
	\right]  \\
& \hspace{-7mm}  
  +  \ExpALL \left\| \PIjjRXavg \rhojjRX \PIjjRXavg   - \rhojjRX \right\|_1 \\
&\leq 
	1 - (1-\epsilon) + 2\sqrt{\epsilon}.
\end{align*}
The %
inequality follows from 
$\ExpUXgXone \!\!\!\left\{ \rhojjRX   \right\} = \tau_{\ell_j}$,
the %
properties of typical projectors \cite{wilde2011book}:
$\ExpXone\text{Tr}[  \PIjjRXlj \ \tau_{\ell_j} ] \geq 1- \eps$,
$\text{Tr}[  \PIjjRXavg \ \bar{\tau} ] \geq 1- \eps$
and Lemma~\ref{lem:gentle-operator}.

The error term $\alpha$ is bounded in a similar fashion.

\medskip
We can split the sum in the second type of error, (II),
as $\sum_{\left(  \ell_{j}^{\prime}%
,m_{j}^{\prime}\right)  \neq\left(  \ell_{j},m_{j}\right)  }\left(  \cdot\right)
=\sum_{m_{j}^{\prime}\neq m_{j}}\left(  \cdot\right)  +\sum_{\ell_{j}^{\prime
}\neq \ell_{j},\ m_{j}^{\prime}}\left(  \cdot\right)  $:
\begin{align*}
& \!\!\!\!\!\!\!
\sum_{\left(  \ell_{j}^{\prime},m_{j}^{\prime}\right)  \neq\left(\ell_{j},m_{j}\right)  }
\!\!\!\!\!
\text{Tr}\!
\left[
	\PRXljprmjpr \ \ \rhoRX
\right] \\
& \ \ \ \ = 
\underbrace{
\sum_{m_{j}^{\prime}\neq m_{j}}
\!\!\!\!
\text{Tr}\!
\left[
	\PRXljmjpr \ \ \rhoRX
\right] 
}_{(A)}
\\
& \qquad \qquad + \!
\underbrace{
\sum_{l_{j}^{\prime}\neq l_{j},\ m_{j}^{\prime}}
\!\!\!\!\!
\text{Tr}\!
\left[ 
	\PRXljprmjpr \ \ \rhoRX
\right]
}_{(B)}.
\end{align*}

We now analyze the two terms $(A)$ and $(B)$ separately.

\paragraph{Matching $\ell_j$, wrong $m_j$} 
	By performing the error analysis for the case 
	where $\ell_j$ is decoded correctly, but $m_j$ is decoded incorrectly,
	we obtain the bound $R_m < I(X;B|UX_1)=H(B|UX_1)-H(B|UXX_1)-\delta$,
	using the following properties of typical projectors \cite{wilde2011book}:
	\begin{align}
	\PIjRXmjprlj \!\!\leq  & 2^{n[H(B|UXX_1)+\delta ]}  \rhojRXmjpr, 
			\label{eqn:for-proj-trick}\\
		  \!\!\!\!\PIjRXlj \! \rhobarRX \PIjRXlj   \!\! \leq & 2^{-n[H(B|UX_1)-\delta] }  \PIjRXlj. \label{eqn:for-proj-trick2}
	\end{align}
Consider the first term:%
\begin{align*}
& \!(A) =   \sum_{m_{j}^{\prime}\neq m_{j}}
\!\!\!
\text{Tr}\!
\left[
	\PRXljmjpr \ \ \rhoRX
\right] \\
& = \!\!\!\!
	\sum_{m_{j}^{\prime}\neq m_{j}}
	\!\!\!\!\text{Tr}\!
	\left[  
	\left(  
	\PRXjmjpr \! \otimes \! \PRXjj
	\right)  
	\rhojRX \!\otimes \!\rhojjRX
	\right]  
	\\
&  \leq \!\!\!\!
	\sum_{m_{j}^{\prime}\neq m_{j}}
	\!\!\!\!\text{Tr} \! \left[  %
	\PRXjmjpr %
	\! \otimes \! I^{B_{\left(  j+1\right)  }^{n}} %
	\ \ 
	\rhojRX \otimes \rhojjRX
	\right]  %
	\\
&  = \!\!\!\!
	\sum_{m_{j}^{\prime}\neq m_{j}	}
	\!\!\!\!\text{Tr}\!
	\left[  
	\PRXjmjpr
	\ \
	\rhojRX
	\right]
	\\[-3mm]
&  =
	\!\!\!\!\!\!
	\sum_{m_{j}^{\prime}\neq m_{j}	}
	\!\!\!\!\!
	\text{Tr}\!
	\Bigg[
	\PIjRXavg
	\!
	\underbrace{
	\PIjRXlj
	\!\!\!
	\overbrace{\PIjRXmjprlj \!\!\!\!\!\!\!\!\!\! }^{\textrm{\dingone}} %
	\ 
	\PIjRXlj
	\!\!
	}_{\textrm{\dingtwo}}
	\PIjRXavg
	\rhojRX \!
	\Bigg]  %
\end{align*}
We now upper bound expression \dingone using \eqref{eqn:for-proj-trick}
and take the conditional expectation with respect to $X^n$:
\[
	\ExpXgUXone\!\!\left\{ \rhojRXmjpr \right\} = \rhobarRX,
\]
which is independent of the state $\rhojRX$ since $m^\prime_j \neq m_j$.
The resulting expression in \dingtwo has the state 
$\rhobarRX$ sandwiched between its
typical projector on both sides, and so we can use \eqref{eqn:for-proj-trick2}.
After these steps, we obtain the upper bound:
\begin{align}
& \!\!\!\! \ExpXgUXone\!\!\left\{  (A) \right\}  
\leq
2^{n\left[   H\left(  B|XUX_{1}\right)  +\delta\right]  }  \ 
2^{-n\left[  H\left(B|UX_{1}\right)  -\delta\right]  } \times \nonumber  \\[-2mm] 
& \quad \quad
\ExpXgUXone
	\!\!\sum_{m_{j}^{\prime}\neq m_{j}	}\!\!\!
	\text{Tr}\!
	\left[
		\PIjRXavg
		 \PIjRXlj
		\PIjRXavg
	\
		\rhojRX 	
	\right]   \nonumber  \\
&  \leq
	2^{n\left[  H\left(  B|XUX_{1}\right)  +\delta\right]  }
	2^{-n\left[  H\left(B|UX_{1}\right)  -\delta\right]  } 
	\sum_{m_{j}^{\prime}\neq m_{j}	} \text{Tr}\!\left[ \ \rhojRX \ \right]
	 \nonumber  \\[-3mm]
& \leq
	|\mcal{M}| \  %
	2^{-n\left[  I\left(  X;B|UX_{1}\right)-2\delta\right]  }
	.
	\label{eqn:bound-on-Rm}
\end{align}
The first inequality follows because each operator
inside the trace is positive and less than the identity.

\paragraph{Wrong $\ell_j$ (and thus wrong $m_j$)}
	We obtain the bound 
	$R \equiv R_\ell + R_m \leq I(XX_1;B)=I(X_1;B)+I(UX;B|X_1)$
	from the ``{\sc and}-measurement'' and the following inequalities:
	\begin{align}
		\Tr[ \PIjjRXlj ] & \leq 2^{n[H(B|X_1)+\delta ]}, \\
		 \PIjjRXavg \  \bar{\tau}  \ \ \PIjjRXavg & \leq 2^{-n[H(B)-\delta] } \PIjjRXavg, \\
		\Tr[ \PIjRXmjlj ] & \leq 2^{n[H(B|UXX_1)+\delta ]},  \label{eqn:rhobar-proj-size}\\
		 \PIjRXavg \  \rhodbarRX  \ \PIjRXavg & \leq 2^{-n[H(B|X_1)-\delta] } \PIjRXavg.
		 \label{eqn:rhodbar-sandwich} 
	\end{align}	

Consider the following term:%
\begin{align*}
\!\!(B) 
&=	\!\!\!\!\!
	\sum_{\ell_{j}^{\prime}\neq \ell_{j},m_{j}^{\prime}}
	\!\!\!\!\!
	\text{Tr}\!
	\left[ 
		\PRXljprmjpr \ \ \rhoRX
	\right] \\
& = \!\!\!\!
	\!\!
	\sum_{\ell_{j}^{\prime}\neq \ell_{j},m_{j}^{\prime}}
	\!\!\!\!\text{Tr}\!
	\left[  
	\left(  
	\PRXjljprmjpr \!\! \otimes \! \PRXjjpr
	\right)  
	\rhoRX
	\right]  
	\\	
& = \!\!\!\!  
	\!\!
	\sum_{\!\! \ell_{j}^{\prime}\neq \ell_{j},m_{j}^{\prime} \ }
	\!\!
	\underbrace{
	\!\!\!\!\text{Tr}\!
	\left[  
		\PRXjljprmjpr 
		\rhojRX
	\right] 
	}_{(B1)}
	\underbrace{
	\!\text{Tr}\!
	\left[  	
		\PRXjjpr
		 \rhojjRX
	\right]  
	}_{(B2)}
	\vspace{-1mm}
\end{align*}

We want to calculate the expectation of $(B)$ under
the code randomness $\ExpALL$.   
The random variables in different blocks are independent, and
so we can analyze the expectations
of the terms $(B1)$ and $(B2)$ separately. 

Consider first the calculation in block $j$, which leads
to the following bound on the expectation of $(B1)$:
\vspace{-1mm}
\begin{align*}
& \ExpALL \!  \left\{ (B1) \right\}  
= 
	\ExpALL \! \left\{ 
	\text{Tr}\!
	\left[  
		\PRXjljprmjpr 
		\rhojRX
	\right] 
	\right\} \\
& = \ExpALL
	\text{Tr}\left[
	\begin{array}[c]{l}%
		\PIjRXljpr
		\PIjRXmjprljpr
		\PIjRXljpr \times \\
		\qquad  \qquad \quad
		\PIjRXavg
		\rhojRX 
		\PIjRXavg
	\end{array}
	\right]
	\\
& = 
\ExpXone
	\text{Tr}\left[
	\begin{array}[c]{l}
		\displaystyle \ExpUXgXone \{
		\PIjRXljpr
		\PIjRXmjprljpr
		\PIjRXljpr \} \times \\
		\qquad  \quad 
		\PIjRXavg
		\underbrace{
		\displaystyle \ExpUXgXone \!\! \left\{ \
		\rhojRX  \
		 \right\}
		 }_{\ \textrm{\dingthree}  }
		  \PIjRXavg
	\end{array}
	\right]
	\\[-1.5mm]
& = 
\ExpXone
	\text{Tr}\left[
	\begin{array}
	[c]{l}%
	\displaystyle \ExpUXgXone \{
	\PIjRXljpr
	\PIjRXmjprljpr
	\PIjRXljpr \} \times \\
	\qquad  \qquad  \qquad
	\underbrace{
	\PIjRXavg
	\rhodbarRX
	 \PIjRXavg
	 }_{\textrm{\ \dingfour}}
	\end{array}
	\right]
	\\	
& \leq 
2^{-n\left[  H\left(  B|X_{1}\right)  -\delta\right]  }
\!\!\!\!\!\ExpALL \!\!\!\!\!\!\!
	\text{Tr}\!\left[ \!\!
	\begin{array}[c]{l}%
	\PIjRXljpr \!\!
	\PIjRXmjprljpr \!\!
	\PIjRXljpr  \!
		\PIjRXavg
	\!\!\!\!
	\end{array}
	\right]
	\\	
& \leq 
2^{-n\left[  H\left(  B|X_{1}\right)  -\delta\right]  }
\!\!\!\!\!\ExpALL \!\!\!
	\text{Tr}\left[
	\PIjRXmjprljpr
	\right]
	\\	
& \leq
	2^{-n\left[  H\left(  B|X_{1}\right)  -\delta\right]  }
	\!\!\!\ExpALL \!\!\!
	2^{n\left[H\left(  B|X_{1}UX\right)  +\delta\right]  }\\
& =
	2^{-n\left[  I\left(  UX;B|X_{1}\right)  -2\delta\right]  }
\vspace{-1mm}
\end{align*}
The result of the expectation in \dingthree is $\rhodbarRX$,
and we can bound the expression in  \dingfour using \eqref{eqn:rhodbar-sandwich}.
The first inequality follows because all the other terms in the trace are positive operators
less than the identity. The final inequality follows from \eqref{eqn:rhobar-proj-size}.

Now we consider the expectation of the second term:
\begin{align*}
 \ExpALL & \!  \left\{ (B2) \right\}  
= 
	\ExpALL
	 \left\{  \text{Tr}\left\{  
	 \PRXjjpr \ 
	\rhojjRX
	\right\}  \right\}  \\
&  =\text{Tr}\left\{  
 \PRXjjpr \ 
\ExpALL
\left\{  
\rhojjRX
\right\}  \right\}  \\
&  =\text{Tr}\left\{  
 \PRXjjpr \ 
\bar{\tau}^{\otimes n}
\right\}  \\
&  =\text{Tr}\left\{  
\PIjjRXavg
\PIjjRXljpr \
\PIjjRXavg \
\bar{\tau}^{\otimes n}
\right\}  \\
&  =\text{Tr}\left\{  
\PIjjRXljpr  \
\PIjjRXavg 
\bar{\tau}^{\otimes n}
\PIjjRXavg
\right\} 
\end{align*}
\begin{align*}
&   \leq
2^{-n\left[  H\left(  B\right)  -\delta\right]  }
\text{Tr}\left\{
\PIjjRXljpr \
\PIjjRXavg
\right\}  \\
&  \leq
	2^{-n\left[  H\left(  B\right)  -\delta\right]  }
	2^{n\left[  H\left(B|X_{1}\right)  +\delta\right]  }
  =
	2^{-n\left[  I\left(  X_{1};B\right)  -2\delta\right]  }.
\end{align*}

Combining the upper bounds on $(B1)$ and $(B2)$  %
gives our final upper bound: %
\begin{align}
& \ExpALL \!  \left\{ (B) \right\}   = 
\ExpALL 
\sum_{\ell_{j}^{\prime}\neq \ell_{j},m_{j}^{\prime}}
(B1) \times (B2) \nonumber \\
& \leq
	\sum_{\ell_{j}^{\prime}\neq \ell_{j},\ m_{j}^{\prime}}
	2^{-n\left[  I\left(  UX;B|X_{1}\right) -2\delta\right]  } 
	\times
	2^{-n\left[  I\left(X_{1};B\right)  -2\delta\right]  }	
	\nonumber  \\
& \leq 
	|\mcal{L}||\mcal{M}| \  2^{-n\left[  I\left(  X_{1};B\right)  +I\left(  UX;B|X_{1}\right)  -4\delta\right]  }.
	\label{eqn:bound-on-R}
\end{align}
By choosing the size of 
message sets to satisfy equations \eqref{eqn:bound-from-relay},
\eqref{eqn:bound-on-Rm} and 
\eqref{eqn:bound-on-R},
the expectation of the average probability of error
becomes arbitrarily small for $n$ sufficiently large.
\QED

\section{Discussion}
	\label{sec:discussion}

	We proved the achievability of the rates given by
	the partial decode and forward inner bound,
	thus extending the study of classical-quantum channels
	to multi-hop scenarios.
	An interesting open question is
	to determine a compress-and-forward strategy for the quantum setting.
	Another avenue for research would be to consider \emph{quantum}
	communication scenarios, and results here might have applications
	 for the design of quantum repeaters \cite{collins2005quantum}.
	I.~Savov acknowledges support from FQRNT and NSERC.
	M.~M.~Wilde acknowledges support from the
	Centre de Recherches Math\'ematiques.

\bibliographystyle{IEEEtran}
\bibliography{qrelay}

\newpage

\appendix

\section{Proof conitnued...}

\subsection{Decoding at the Relay}

In this section we give the details of the POVM construction
and the error analysis for the Relay decoder.

\medskip
\textbf{POVM\ Construction}. 
	During block $j$, the Relay wants to decode the 
	message $\ell_j$ encoded in $u^n(\ell_j,\ell_{j-1})$,
	given the knowledge of the message $\ell_{j-1}$ from the previous block.
	Consider the state obtained by tracing over the systems $X$
	and $B$ in (\ref{eq:code-state}):
	\[
	\theta^{UX_1B_1} =
	\sum_{u,x_{1}}p\left(  u|x_{1}\right)  p\left(  x_{1}\right)  \left\vert
	u\right\rangle \left\langle u\right\vert ^{U}\otimes\left\vert x_{1}%
	\right\rangle \left\langle x_{1}\right\vert ^{X_{1}}\otimes\sigma_{u,x_{1}%
	}^{B_{1}},
	\]
	where $
	\sigma_{u,x_{1}}^{B_{1}}\equiv\sum_{x}p\left(  x|x_{1},u\right)  \text{Tr}%
	_{B}\left\{  \rho_{x,x_{1}}^{B_{1}B}\right\}$.
	Further tracing over the system $U$ leads to the state%
	\[
	\theta^{X_1B_1} =
	\sum_{x_{1}}p\left(  x_{1}\right)  \left\vert x_{1}\right\rangle \left\langle
	x_{1}\right\vert ^{X_{1}}\otimes\bar{\sigma}_{x_{1}}^{B_{1}},
	\]
	where $
	\bar{\sigma}_{x_{1}}
	\equiv\sum_{u}p\left(  u|x_{1}\right)  \sigma_{u,x_{1}%
	}^{B_{1}}$.
	Corresponding to the above conditional states are conditionally typical
	projectors of the following form%
	\begin{align*}
	\PIRElj 
	& \equiv
	  \Pi_{\sigma_{u^{n}\left(  \ell_{j},\ell_{j-1}\right)  ,x_{1}^{n}\left(
	\ell_{j-1}\right)  }}^{B_{1\left(  j\right)  }^{n}}, \ \  %
	\PIREavg 
	\equiv 
	 \Pi_{\bar{\sigma}_{x_{1}^{n}\left(  \ell_{j-1}\right)  }}^{B_{1\left(  j\right)
	}^{n}}.
	\end{align*}
	The Relay constructs a square-root measurement $\{ \Gamma_{\ell_j} \}$
	using formula \eqref{eq:square-root-POVM-generic} and the following positive operators:
	\be
	\PRE \equiv
	\PIREavg \PIRElj \PIREavg
	\ee

\medskip
\textbf{Error analysis}. 
	In this section we show that during block $j$ the Relay will be able to decode $\ell_{j}$ 
	from the state $\rhoFULLatRE$,
	provided 
	the rate $R_\ell < I(U;B_1|X_1)=H(B_1|X_1)-H(B_1|UX_1)-\delta$.
	The bound follows from the following properties of typical projectors:
	\begin{align}
		\Tr[ \PIRElj ] &\leq 2^{n[H(B_1|UX_1)+\delta ]}  \label{RE-typ-1}\\
		 \PIREavg  \bar{\sigma}  \ \PIREavg & \leq 2^{-n[H(B_1|X_1)-\delta] } \PIREavg, \label{RE-typ-2}.
	\end{align}

	The average probability of error at the Relay is given by:
	\begin{align*}
		\bar{p}_{e}^{R}
		\equiv 
			\frac{1}{|\mathcal{L}|}\sum_{\ell_j}					
			\text{Tr}\!
			\left\{  
				\left(  I-\Gamma^{B_{1(j)}^{n}} _{\ell_j|\ell_{j-1}} \right)
				\rhojRE
			\right\},
	\end{align*}

	We consider the probability of error for a single message $\ell_j$
	and begin by applying the Hayashi-Nagaoka operator inequality
	(Lemma~\ref{lem:HN-inequality}) to split the error
	into two terms:
	\begin{align*}
	 \bar{p}_e^{R}
	 & \equiv
	\text{Tr}\!\left[ 
		\left(I   - \!\GAMRE \right)  
		\rhojRE
	  \right] \\
	& \leq  2
	\underbrace{
	\text{Tr}\left[  \left(  I 
	-\PRE
	\right)   \ 
	\rhojRE 
	\right]  
	}_{(I)}
	\\
	&  \qquad +4 \!
	\underbrace{
	\sum_{\ell_{j}^{\prime}   \neq \ell_{j}  }
	\!\!
	\text{Tr}
	\left[  
	\PREpr
	\
	\rhojRE
	\right]
	}_{(II)}
	\!.
	\end{align*}
	
	We will bound the expectation %
	of the average probability of error by bounding the 
	individual terms.
	We bound the first term as follows:
	\begin{align*}
	(\text{I})  
	&=	
		\text{Tr}\left[  
			\left(  I  -\PRE\right)   \ 
			\rhojRE 
		\right] \\
	&=
		\text{Tr}\!\left[  
			\left(  I- \PIREavg \PIRElj \PIREavg \right) 
			\rhojRE  
		\right]  \\
	&=
		1 - 
		\text{Tr}\!\left[  
			\PIREavg \PIRElj \PIREavg \ 
			\rhojRE  
		\right]  \\
	&\leq
		1 -
		\text{Tr}\!\left[  
		  \PIRElj \ 
			\rhojRE  
		\right]  \\
	&\qquad \quad 
	  +  \left\| \PIREavg \rhojRE \PIREavg   - \rhojRE \right\|_1,
	\end{align*}
	where the inequality follows from Lemma~\ref{lem:tr-trick}.

	By taking the expectation over the code randomness
	we obtain the bound
	\begin{align*}
	& \!\!\!\ExpALL   \!\!\!\!(\text{I})  
	=	
		1 - 
		\ExpUXone
		\text{Tr}\!\left[  
		  \PIRElj \! 
			\ExpXgUXone \!\!\!\left\{ \rhojRE   \right\}
		\right]  \\
	& \ \ \ \ %
	  +   \!\!\ExpALL \left\| \PIREavg \rhojRE \PIREavg   - \rhojRE \right\|_1 \\
	& =
		1 - 
		\ExpUXone
		\text{Tr}\!\left[  
		  \PIRElj \! 
		  	\sigma_{\ell_j,\ell_{j-1}}
		\right]  \\
	& \ \ \ \ %
	  +  \!\! \ExpALL \left\| \PIREavg \rhojRE \PIREavg   - \rhojRE \right\|_1 \\
	& \leq
		1 - 
		\ExpUXone
		\text{Tr}\!\left[  
		  \PIRElj \! 
		  	\sigma_{\ell_j,\ell_{j-1}}
		\right]  + 2 \sqrt{\epsilon}  \\
	&\leq 
		1 - (1-\epsilon) + 2\sqrt{\epsilon} = \epsilon + 2\sqrt{\epsilon}.
	\end{align*}
	The first inequality follows from Lemma~\ref{lem:gentle-operator}
	and the property
	\be
		\text{Tr}\!\left[  \PIREavg \ \bar{\sigma} \right] \geq 1- \eps.
	\ee
	The second inequality follows from:
	\be
		\text{Tr}\!\left[ 	
			\PIRElj  \sigma_{\ell_j,\ell_{j-1}} 
		 \right] 
		 \geq 1- \eps.
	\ee
	
	\medskip
	To bound the second term we proceed as follows:
	\begin{align*}
	 \ExpALL & \!  \left\{ (II) \right\}  
	 = 
		\ExpALL \sum_{\ell_{j}^{\prime}   \neq \ell_{j}  }
		\!\!
		\text{Tr}
		\left[  
		\PREpr
		\
		\rhojRE
		\right] \\
	 & = \!\!
		\ExpXone \sum_{\ell_{j}^{\prime}   \neq \ell_{j}  }
		\!\!
		\text{Tr}
		\left[  
		\ExpUXgXone \!\!\!\left\{
		\PREpr
		\right\}
		\!\!
		\ExpUXgXone \{
		\rhojRE
		\}
		\right] \\
	& = \!\!
		\ExpXone
		\sum_{\ell_{j}^{\prime}   \neq \ell_{j}  }
		\!\!
		\text{Tr}
		\left[  
		\ExpUXgXone \!\!\!\left\{
		\PREpr
		\right\}
		\ 
		\bar{\sigma}_{|\ell_{j-1}} %
		\right]
	\end{align*}		
	The expectation can be broken
	up because $\ell_j^\prime \neq \ell_j$
	and thus the $U^n$ codewords are
	independent.
	We have also used 
	\be
		\ExpUXgXone \{
		\rhojRE
		\}
		=
		\bar{\sigma}_{|\ell_{j-1}}.
	\ee

	We continue by expanding the operator
	$\PREpr$ as follows:
	\begin{align*}		
	 \ \ 
	 & = \!\!\!\!
		\ExpALL
		\sum_{\ell_{j}^{\prime}   \neq \ell_{j}  }
		\!\!
		\text{Tr}
		\left[  
		\PIREavg \!\!\PIREljpr \!\!\PIREavg
		\bar{\sigma}_{|\ell_{j-1}} %
		\right] \\
	 & = \!\!
		\ExpALL
		\sum_{\ell_{j}^{\prime}   \neq \ell_{j}  }
		\!\!
		\text{Tr}
		\left[  
		\qquad  \PIREljpr 
		\underbrace{
		\PIREavg
		\bar{\sigma}_{|\ell_{j-1}}
		\PIREavg
		}_{\textrm{\dingfive}}
		\right] \\
	 & \leq \!\!
		\ExpALL
		\sum_{\ell_{j}^{\prime}   \neq \ell_{j}  }
		\!\!
		\text{Tr}
		\left[  
		 \PIREljpr  \
 	 	2^{-n[H(B_1|X_1)-\delta] }
		\PIREavg
		\right] \\
	 & \leq 
	 	2^{-n[H(B_1|X_1)-\delta] }
		\ExpALL
		\sum_{\ell_{j}^{\prime}   \neq \ell_{j}  }
		\!\!
		\text{Tr}
		\left[  
		 \PIREljpr   
		\right] \\
	 & \leq 
	 	2^{-n[H(B_1|X_1)-\delta] }
		\ExpALL
		\sum_{\ell_{j}^{\prime}   \neq \ell_{j}  }
		\!\!
		2^{n[H(B_1|UX_1)+\delta ]} \\
	 & \leq 
	 	| \mcal{L}| \
	 	2^{-n[I(U;B_1|X_1)-2\delta] }.
	\end{align*}
	The first inequality follows from using \eqref{RE-typ-2} on the expression \dingfive \!\!.
	The second inequality follows from the fact that $\PIREavg$ is a positive
	operator less than the identity. More precisely we have
	\begin{align*}
		\PIREljpr \PIREavg 
		& = 
			 \PIREljpr \PIREavg \PIREljpr  \\
		& \leq 
			\PIREljpr  I \ \PIREljpr \\
		& = 
			\PIREljpr.
	\end{align*}
	The penultimate inequality follows from
	 \eqref{RE-typ-1}.

	Thus if we choose
	$R_\ell \leq I(U;B_1|X_1)-3\delta$,
	we can make the expectation 
	of the average probability of error 
	vanish in the limit of
	many uses of the channel.

\medskip
\textbf{Proof conclusion}.
Note that the gentle operator lemma for ensembles is used
several times in the proof to guarantee that the effect of
acting with one of the projectors from the ``measurement sandwich'' 
does not disturb the state too much.
Furthermore, because each of the outputs blocks is operated
on twice, we again depend on the gentle operator lemma 
to guarantee that the state disturbance is asymptotically negligible.

\end{document}